\documentclass{iaus}
\usepackage{graphicx}

\title %% give here short title %%
{New Results from High Energy Gamma-Ray Astronomy}

\author   %% give here short author list %%
{H. J. V\"olk}
\affiliation{Max-Planck-Institut f\"ur Kernphysik, P. O. Box, 69029 
             Heidelberg, Germany 
\break email: Heinrich.Voelk@mpi-hd.mpg.de

}

\pubyear{2005}
\volume{230}  %% insert here IAU Symposium No.
\pagerange{1-8}
\date{?? and in revised form ??}
\jname{Proceedings Title IAU Symposium}
\editors{E.J.A. Meurs, G. Fabbiano, eds.}
\begin{document}
    
\maketitle

\begin{abstract}
High energy gamma-ray astronomy has recently made significant progresss through
ground-based instruments like the {\it H.E.S.S.} array of imaging atmospheric
Cherenkov telescopes. The unprecedented angular resolution and the large field
of view has allowed to spatially resolve for the first time the morphology of
gamma-ray sources in the TeV energy range. The experimental technique is
described and the types of sources detected and still expected are
discussed. Selected results include objects as different as a Galactic binary
Pulsar, the Galactic Center and Supernova Remnants but they also concern the
diffuse extragalactic optical/infrared radiation field. Finally, a scan of the
Galactic plane in TeV gamma rays is described which has led to a significant
number of new TeV sources, many of which are still unidentified in other
wavelengths. The field has a close connection with X-ray astronomy which allows
the study of the synchrotron emission from these very high energy sources.
\keywords{radiation mechanisms: nonthermal, pulsars, Galaxy: center, acceleration of
 particles, supernova remnants, gamma rays: observations, cosmology: observations.}
\end{abstract}

%  \section{Introduction} 
% if your document starts with a section,
% Recently gamma-ray astronomy at very high energies (VHE) has made great
% progress through the detection of many new sources and through detailed
% theoretical efforts regarding the physics of the nonthermal component in
% astrophysical objects. The observational results have been obtained with
% ground-based imaging Air Cherenkov Telescopes. I will first summarize the
 % type s
% of sources expected and indicate the detections which include a recent
% investigation of the diffuse extragalactic radiation field due to stars and
% dust. Then I will discuss a number of specific objects and their astrophysical
% interpretation. These are the Binary Pulsar PSR B1259-630, the Galactic
% Center ,
% and several recently observed Supernova Remnants. Finally I will report on the
% recent Galactic Plane Scan achieved with the {\it H.E.S.S.} experiment.

\begin{figure}
\includegraphics[width=5.3in,angle=0]{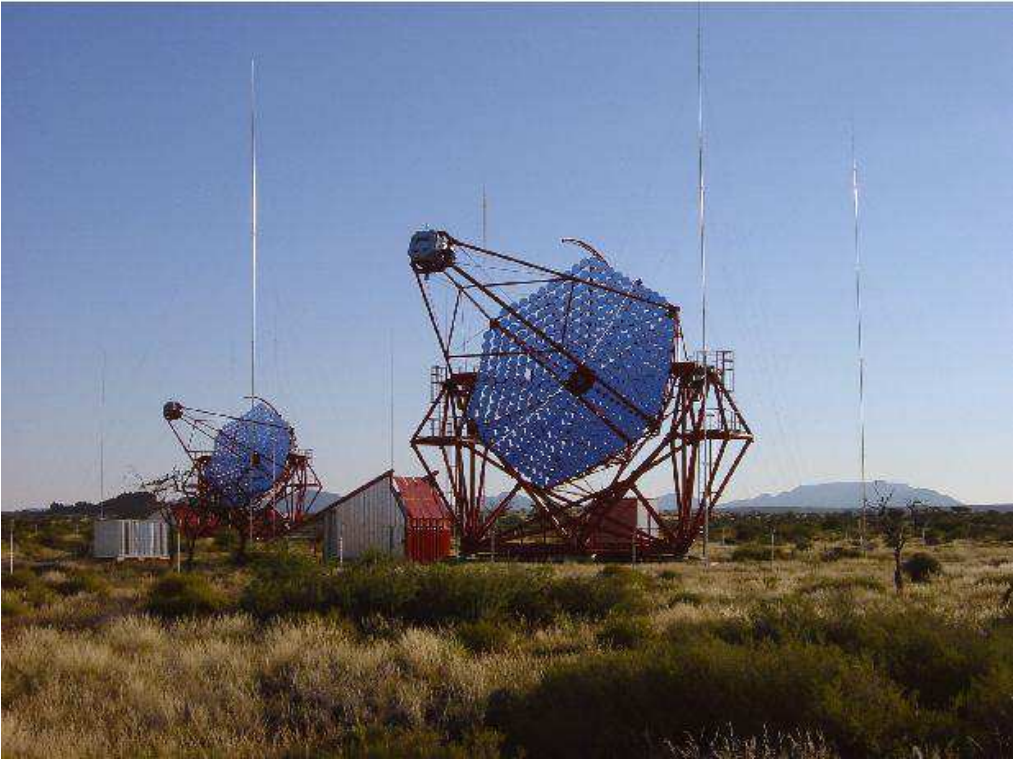}
  \caption{The {\it H.E.S.S.} experiment in Namibia, near the famous
    Gamsberg. Two of the four 13m-telescopes are shown with their ``cameras''
    in the focal point.}
\label{fig:HESS}
\end{figure}

\section{Imaging Air Cherenkov Telescopes and the H.E.S.S. experiment}

The principles of the Air Cherenkov Technique were invented in the 1950ies,
with a first astronomical success in the late 1980ies, when the {\it Whipple}
telescope in Arizona detected the Crab Nebula (\cite{Weekes89}).

The recent progress in VHE $\gamma$-ray astronomy with the {\it H.E.S.S.}
experiment in Namibia was possible through several new developments in the
1990ies (i) the stereoscopic technique, pioneered by the
German/Spanish/Armenian {\it HEGRA} experiment on La Palma (ii) the development
of fine pixel ``cameras'' as focal plane detectors ({\it CAT} telescope in the
French Pyrenees) (iii) the realization of the spatial extension of most of the
expected $\gamma$-ray sources and thus the necessity for wide field-of-view
(FoV) detectors allowing also sky surveys, and (iv) the need for the detection
of ``many'', spatially extended nearby sources to solve the problem of Cosmic
Ray origin in our Galaxy. It is therefore advantageous to observe the Galactic
disk in the southern sky. This led to the choice of the Gamsberg area in
Namibia in Southern Africa.

Fig.1 shows two of the four telescopes of the {\it High Energy Stereoscopic
System (H.E.S.S.)} at 1800 m.a.s.l. It is operated by a large collaboration led
by the Max Planck Institute for Nuclear Physics in Heidelberg
(http://www.mpi-hd.mpg.de/hfm/HESS/HESS/html). The four telescopes operate in
coincidence, at the corners of a square with 100 meters side lengths, just
inside the Cherenkov light cone of an electromagnetic shower of a few ns
duration. It is produced by a high-energy cosmic $\gamma$-ray hitting the
atmosphere and exhibits the maximum of the secondary  electron/positron density
at a height of about 10 km .

Such a stereoscopic device allows - in the manner of a land surveyor - a purely
geometric reconstruction of the direction of the primary photon. Also the
primary energy can be determined through a determination of the shower
footpoint on the ground (\cite{Aharon97}). Gamma-ray events are distinguished
from the much more frequent isotropic events due to charged Cosmic Rays through
their more slender shower images in the cameras, i.e. by image analysis and the
resulting cuts on the image parameters. This permits $\gamma$-ray astronomy at
TeV energies (1 TeV = $10^{12}$~eV)

% \begin{figure}
%  \includegraphics{fig02.eps}
%   \caption{Schematic of the stereoscopic observation of optical Cherenkov
%     emission by the secondary electrons in the atmospheric shower of a cosmic
%     $\gamma$-ray. The Earth's atmosphere is therefore part of the detector
%     an the combined images from the cameras allow the determination of the
%     direction and energy of the primary $\gamma$-ray.}\label{fig:Stereo}
% \end{figure}
%

The characteristics of the {\it H.E.S.S.} telescope system are as follows: 4
telescopes with 107 $m^2$ mirror surface, 4 ``smart'' cameras, each with 960
photomultiplyer tubes with $(0.16^{\circ})$ FoV, leading to a total FoV of~
$\approx 5^{\circ}$ per camera. The corresponding angular resolution of the
system is better than 0.1 \% per event, the energy resolution is about 10 to
15\% per event, and the energy threshold (at Zenith) is about 100 GeV. The
system is fully operational since December 2003.

The resulting {\it H.E.S.S.} sensitivity (4 telescopes) is as follows: 1 hour
of observation time for a detection of an energy flux density of
$10^{-11}~(10^{-12})$ erg cm$^{-2}$ s$^{-1}$ at 100 GeV~(1 TeV). With this
performance the Crab Nebula can be detected at Zenith in $\sim 30$~s. For
comparison, the 1989 detection required $\sim 50$~hr.

\section{Science with high energy gamma-ray astronomy}

The two main fields of very high energy (VHE) $\gamma$-ray astronomy are High
Energy Astrophysics and Observational Cosmology.

\subsection{High Energy Astrophysics}

High Energy Astrophysics concerns the most energetic and violent processes in
the Universe, and in particular their nonthermal aspects. We have to expect
that the nonthermal energy content $U_{nonth}$ of relativistic baryonic
particles is in most regions of the Universe comparable to the energy densities
in the thermal gas $U_{th}$ and the magnetic fields $U_{mag}$, i.e. $U_{nonth}
\sim U_{th} \sim U_{mag}$. This should at least be true ``everywhere'' in
galaxies and clusters of galaxies. But it holds probably also beyond, wherever
cosmic structure formation with its violent, supersonic flows of baryonic
matter has taken place or is still occurring. As a result of interactions of
individual particles with collective excitations of the system, the particle
sources and the associated nonthermal radiation should be characterized by
power-law energy spectra rather than by thermal Maxwellian distributions
 
Because of its expected ubiquity I call this component the ``Nonthermal
Universe''. Its study is intimately connected with that of stellar explosions,
rapid outflows from galaxies, energy losses of extreme compact objects, and
high-energy accretion processes up to the very largest spatial scales.

% \begin{figure}
%  \includegraphics{fig03.eps}
%   \caption{The energy distribution of a gas in which diffusive particle
%     acceleration due to a shock has generated a nonthermal plasma component 
%     besides the heated thermal plamsa. Suprathermal
%    particles with energies $E$ significantly larger than $E_{th}$ start to be
%    accelerated (``injected'') into a power law distribution whose energy can
%    extend over many decades to contain a comparable amount of energy despite
%    the small number density.}
%    \label{fig:distributions}
%  \end{figure}

The types of VHE $\gamma$-ray sources found/{\it expected} in the Galaxy are
Pulsar Nebulae, Supernova Remnants, X-ray Binaries (``Micro-Quasars''), Diffuse
Galactic emission, {\it Molecular Clouds}, or possibly {\it new source
types}. Extragalactic sources are  Active Galactic Nuclei (e.g. Blazars), Radio
Galaxies, {\it Gamma-ray Pair Halos, Starburst Galaxies, Galaxy Mergers, 
Clusters of Galaxies, and Gamma-Ray Bursts.}

\subsection{Observational Cosmology}

A major aspect of cosmology is cosmic structure formation. One of its
consequences is the Extragalactic Background Light (EBL), the thermal, diffuse
optical/infrared extragalactic background radiation from stars and Black Holes
in galaxies and its re-radiation in the infrared. The EBL informs us about the
epochs of galaxy formation and the history of their evolution.

TeV $\gamma$-quanta are absorbed by pair production on these low energy photons
and therefore the spectra of distant extragalactic VHE $\gamma$-ray sources are
expected to exhibit characteristic absorption features. They are the result of
the magnitude and spectral variation of this background. Whereas a direct
astronomical measurement of the EBL is very difficult, the $\gamma
\gamma$-absorption is free of such complications. Its measurement for objects
at different redshifts $z$ should in principle allow even the resolution of the
EBL in $z$. Very recently measurements of two different Blazars with known
redshift that were newly discovered by the {\it H.E.S.S.} Collaboration have
indeed been used (\cite{Aharon05a}; see also the talk by J. Quinn in these
Proceedings) to give the most stringent upper limit on the EBL in the
optical/near-infrared band to date. It appears significantly lower than
expected from the current ``direct'' estimates and very close to the absolute
lower limit represented by the integrated light of resolved galaxies
(\cite{Madau00}). Apart from resolving the EBL largely into individual
galaxies, this upper limit is especially in conflict with the claims of a high
EBL flux at near-infrared wavelengths (\cite{Matsumoto05}; \cite{Cambresy01})
which was often envisaged to be the result of radiation massively produced in
the early Universe ($z \sim 10$) by the first stars (Pop III). Given the strong
likelyhood that Pop III stars rapidly enrich their environment with heavy
elements and dust grains, such an assumption was in any case highly
problematic.

Another major cosmological aspect of VHE $\gamma$-ray astronomy is an indirect
Dark Matter search through the detection of annihilation radiation from the
lowest-mass supersymmetric particles, called Neutralinos. Such weakly
interacting massive particles are widely believed to constitute the
non-baryonic Dark Matter in the Universe.  From gravitational simulations
(e.g. \cite{Navarro96}) they should be concentrated with a significant density
increase in the central regions of Dark Matter halos, like the Galactic center
(see below).

\section{Selected results}

\begin{figure}
\includegraphics[width=5.3in,angle=0]{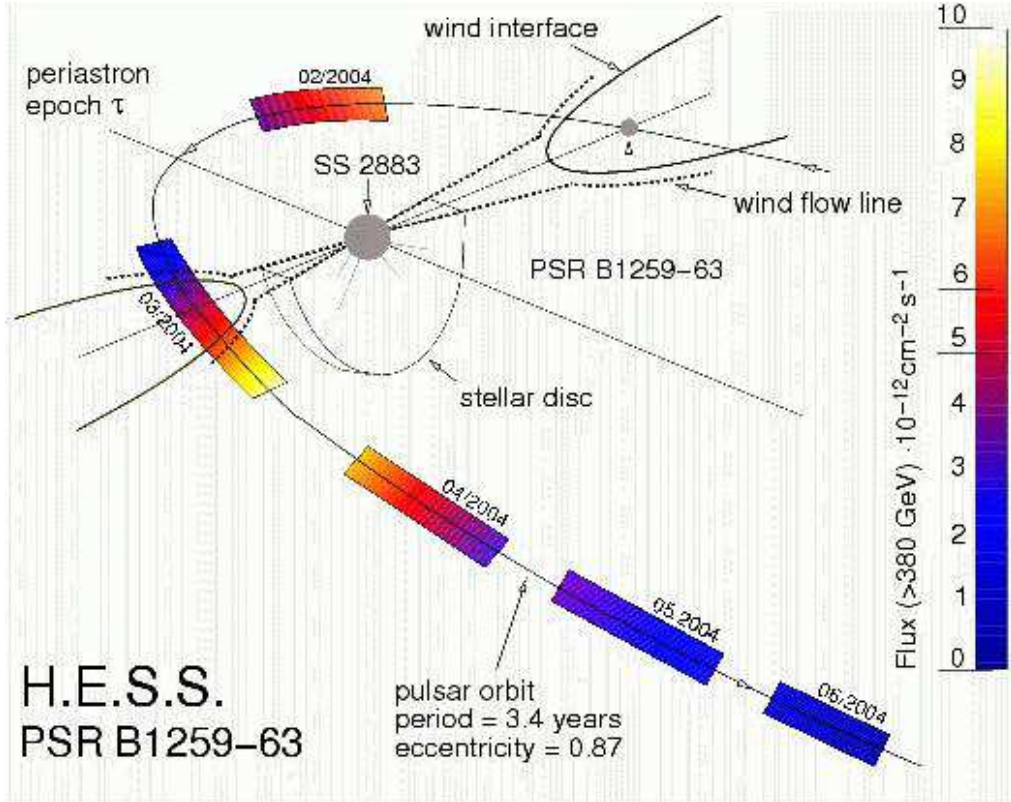}
   \caption{Orbital scheme of PSR B1259-63 with respect to the line of sight 
     (adapted from (\cite{Johnston99}). The Pulsar approaches the stellar wind
      equatorial plane prior to periastron ``behind'' it.}
         \label{fig:Pulsarorbit}
\end{figure}

\subsection{{\it H.E.S.S.} discovery of Pulsar B1259-630}

This 48 ms Pulsar is in a 3.4 yr-period highly excentric orbit around a Be star
with blackbody temperature $T_{*} = 23.000$~K. The particles from the Pulsar
Wind are expected to be ulrarelativistic electrons and positrons, emitting
Inverse Compton (IC) radiation in the stellar radiation field.
Indeed the TeV-spectrum has been successfully predicted in the synchrotron/IC
context (\cite{Kirk99}). {\it H.E.S.S.} has observed the source for $\sim
50$~hrs in 2004 around perihelion; the flux was time-variable on the scales of
days (\cite{Aharon05b}).
%
%  Surprisingly, a second and steady VHE source was found
% in the same FoV. It is the unidentified source HESS J1303-631 with a rather
% hard photon spectrum $\propto E^{-2.2 \pm 0.2}$ (\cite{Aharon05c}). No unique
% counterpart has been found until now.

The Be star emits a strong disk-shaped wind that confines the Pulsar Wind
Nebula into a ``cometary'' shape (where the analogy is meant morphologically
rather than physically). The periastron lies behind the star
(Fig.3). Unfortunately, moonlight did not allow observations during
periastron itself. The data appear nevertheless to indicate a minimum
of the VHE light curve near periastron (Fig.3).

\begin{figure}
\includegraphics[width=5.3in,angle=0]{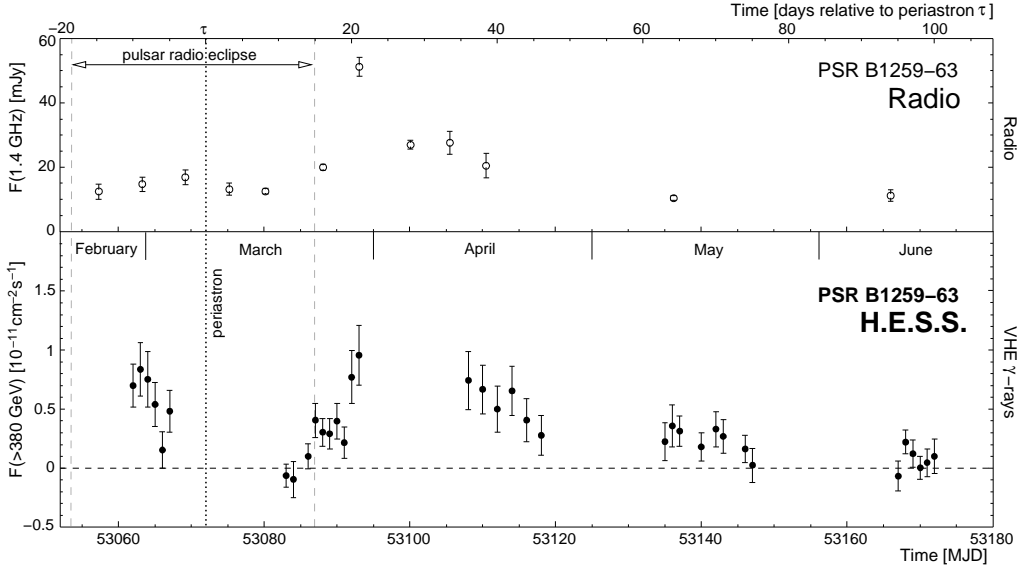}
  \caption{VHE $\gamma$-ray and unpulsed
           radio (\cite{Johnston05}) light curves from PSR B1259-63 around its 
           periastron passage ({\it dotted vertical line}).}
            \label{fig:Pulsarlightcurve}
\end{figure}

The IC losses, as measured by {\it H.E.S.S.}, are about a factor of 10 weaker
than the corresponding synchrotron losses, as inferred by {\it INTEGRAL}
(\cite{Shaw04}). This determines an effective magnetic field strength
$B_{\mbox{eff}}\approx 1$~G. In the expected Klein-Nishina regime the observed
IC photon spectrum corresponds to a very hard differential electron spectrum
with a power-law index $\alpha \geq 1.7 \pm 0.2$. It therefore cannot in turn
come from a radiatively cooled electron population, and adiabatic losses must
be dominant. This points to a rapidly and directionally expanding Pulsar Wind
Nebula flow, resulting in a very interesting physical picture: The shocked
Pulsar Wind is accelerated into a ``cometary tail'' flow, induced by the
stellar wind. This flow is more strongly expanding outside the wind ``disk''
near periastron, where it is directed away from us. As a consequence its
emission is in addition diminished by Doppler dis-favoritism. Both effects
together lead to a local minimum of the IC flux around periastron, despite the
fact that the stellar radiation field peaks at this point.

\subsection{Galactic Center region}

The innermost region of our Galaxy has an abundance of nonthermal sources, many
of them Supernova Remnants. {\it H.E.S.S.} has observed this region and has
detected the neighborhood of Sgr A$^{*}$ with high significance
(\cite{Aharon04a}). Two discrete sources were found, one around Sgr A$^{*}$
itself and another one about $1^{\circ}$ away in the Galactic Disk (see
Fig.4a). This latter source is the Supernova Remnant (SNR) G0.9+0.1, dominated
by a Pulsar Wind Nebula (\cite{Aharon05d}. Although the statistical angular
resolution of the {\it H.E.S.S.} observation of the Sgr A$^{*}$ region is more
than an order of magnitude better than the one achieved in the previous
detections at TeV energies (\cite{Tsuchya04}; \cite{Kosack04}) or in the GeV
range (\cite{Mayer-H98}), the {\it H.E.S.S.} image center cannot yet be
seperated from Sgr A$^{*}$ itself within the present pointing accuracy of the
system ($\sim 20$~arcsec in both coordinates). The $\gamma$-ray source is
apparently point-like.  Therefore other properties like the energy spectrum or
a possible time variability become important criteria to determine whether the
$\gamma$-ray emission comes from the central Black Hole and/or its immediate
environment (the accretion flow or jets), or from a more extended source in the
neighborhood. No time variability has been found until now. The form of the
differential $\gamma$-ray spectrum $\propto E^{-2.3}$ (Fig. 4b) corresponds to
a relatively hard power-law, and is therefore also consistent with freshly
produced Cosmic Rays interacting with dense gas (with a hydrogen density $N_H
\sim 10^3 \mbox{cm}^{-3}$), or diffusive shock acceleration of particles still
confined in the adjacent SNR Sgr A East. Theoretically there is even the
possibility of a continuous transition between these two last radiation
mechanisms.

\begin{figure}
\hbox{
\includegraphics[width=2.6in,angle=0]{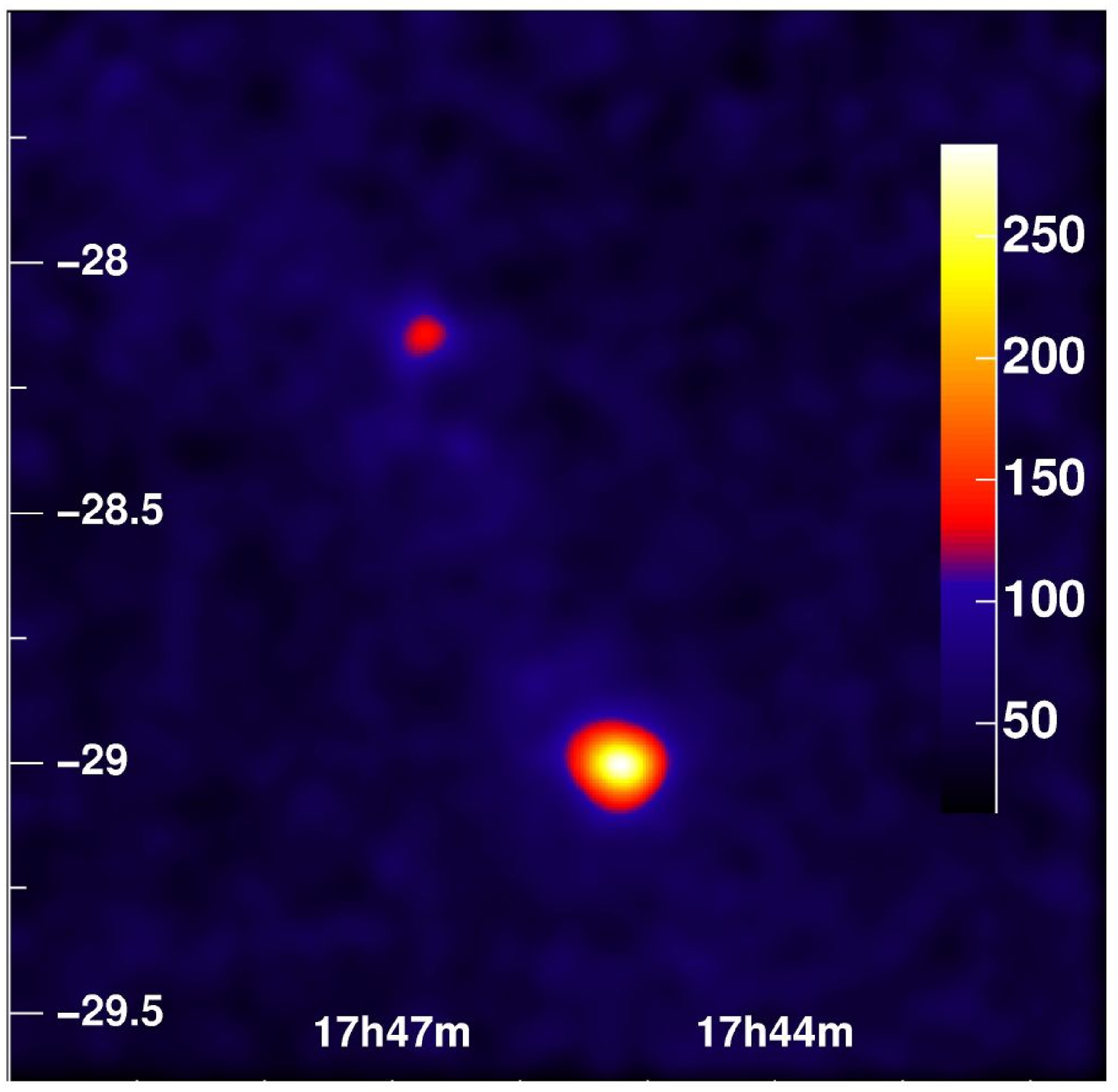}
\includegraphics[width=2.6in,angle=0]{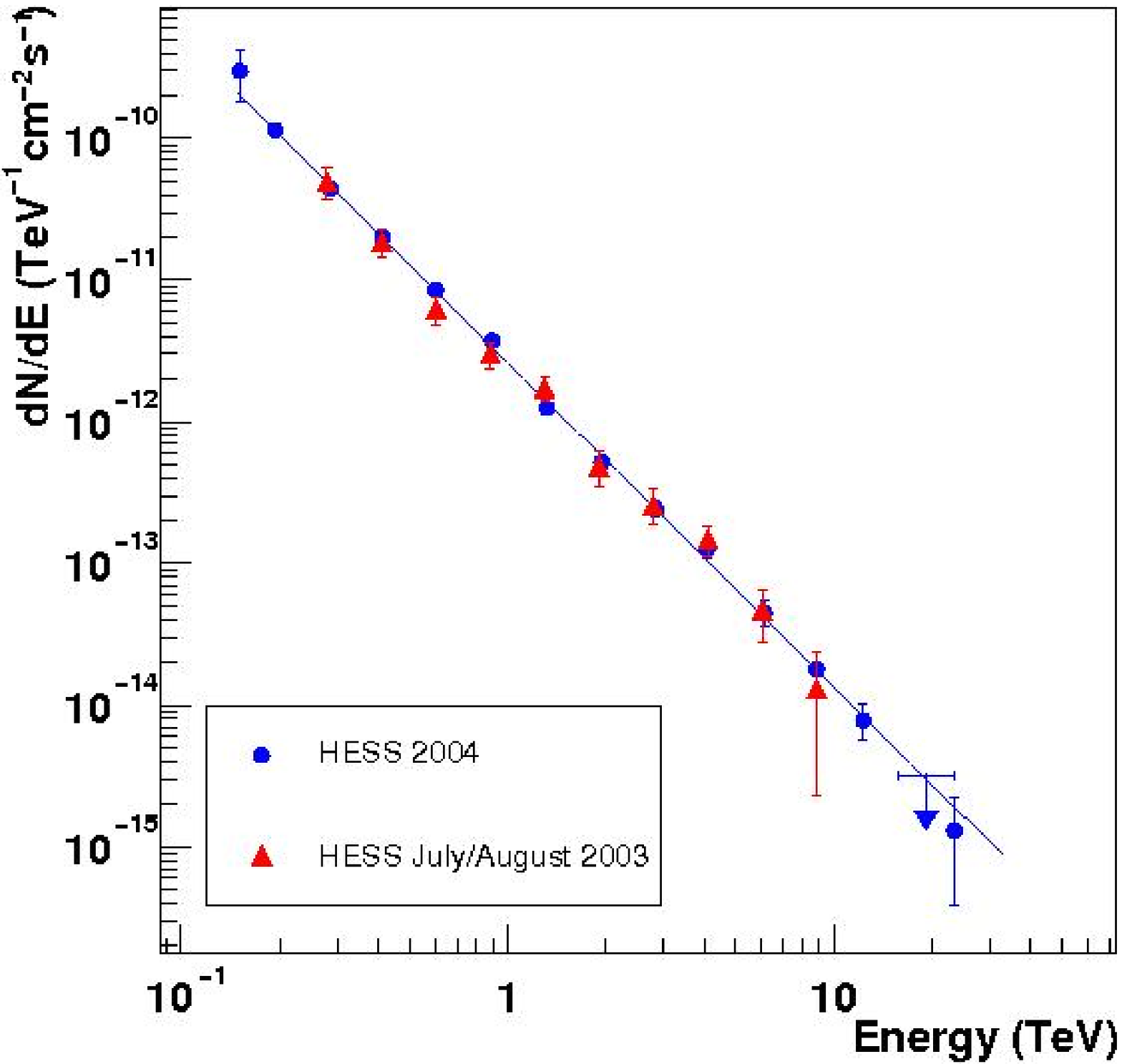}
}
  \caption{(Left:) {\it H.E.S.S.} VHE $\gamma$-ray image of the inner region of the
    Galaxy. The source on the lower right ({\it Galactic Center region}) is
    coincident within $1^{\prime}$ of Sgr A$^{*}$. The source on the upper left
    is the Supernova Remnant G0.9+0.1 in the Galactic plane. (Right:) VHE
    $\gamma$-ray spectrum of the {\it Galactic Center region} as obtained by
    the {\it H.E.S.S.} array. The energy spectrum is a power law and extends
    beyond 20 TeV.}
  \label{fig:GC_image}
\end{figure}
%

%\begin{figure}
% \includegraphics{fig04b.eps}
%\includegraphics[height=2.5in,width=3in,angle=0]{fig04b.eps}
%  \caption{VHE $\gamma$-ray spectrum of the Galactic Center region as obtained
%    by the {\it H.E.S.S.} array. The energy spectrum is a power law and extends
%    beyond 20 TeV.}
%       \label{fig:GC_spectrum}
%\end{figure}
%

The most exotic scenario would be dominant steady-state Neutralino/WIMP
annihilation in the expected Dark Matter Halo centered on Sgr A*. Apart from
the present difficulties to model the observed TeV spectrum within
supersymmetric theories, the rest energy of the Neutralino must then be larger
than the maximum energy up to which the supposed annihilation spectrum extends.
This corresponds to at least 20 TeV. Since the largest particle accelerator to
exist in the foreseeable future, the {\it LHC} in CERN, only reaches
center-of-momentum energies of the order of a TeV, this particle could not
possibly be discovered in accelerator experiments. Thus, if the detected
$\gamma$-ray flux from the Galactic center is to be ascribed to Dark Matter
annihilations, then only astrophysics can tackle this question -- a very
interesting perspective. 

% Presumably
% only the detection of a spectral line feature (e.g. \cite{Bergstrom00}, and
% references therein) could discriminate the annihilation process against the
% other, less speculative emission processes mentioned above. Therefore, the
% Galactic Center region remains one of the most interesting places to be
% observed in VHE $\gamma$-rays, for a number of reasons.

\subsection{Supernova Remnants and Cosmic Ray origin}

% This is a long-standing hypothesis since (\cite{Baade34}}
% and its theoretical basis has been significantly broadened over the years
% (e.g. \cite{V\"olk04} for a recent overview ). However, Cosmic Ray production
% at the outer SNR shock is up to now still only a popular belief and, as in
% other areas of physics, there is the obvious need for agreement between
% experiment and theory.

While the theoretical aspects are essentially understood (e.g. \cite{Voelk04}
for a recent overview), there is a shortage of $\gamma$-ray detections. The
{\it EGRET} experiment could not unequivocally establish the detection of a
shell-type SNR in the GeV-range. Also the widely publicised TeV $\gamma$-ray
detection of SN 1006 by the {\it CANGAROO} experiment could not be confirmed by
{\it H.E.S.S.}  (\cite{Aharon05e}). The reason is in all probability the very
low gas density $N_H < 0.1 \mbox{cm}^{-3}$ (\cite{Ksenofontov05}) in this
object that is so bright in nonthermal X-rays (\cite{Koyama95}). The second
reason for the low TeV flux is the amplification of the magnetic field
({\cite{Bell01}, \cite{Berezhko03a}) which depresses the IC emission -- for
  given synchrotron emission -- compared to that in a more typical $\mu$G
  interstellar field strength. Meanwhile also the {\it CANGAROO} collaboration
  has announced that it could not any more detect SN 1006 in stereoscopic
  re-observations (\cite{Mori05}).

There are only three VHE detections until now:  RX J1713.7-3946
(\cite{Muraishi00}; \cite{Enomoto02}; \cite{Aharon04b}), Cas~A
(\cite{Aharon01}), and  RX J0852.0-4622 (\cite{Katagiri05}; \cite{Aharon05f}).

\begin{figure}
\includegraphics[width=5.3in,angle=0]{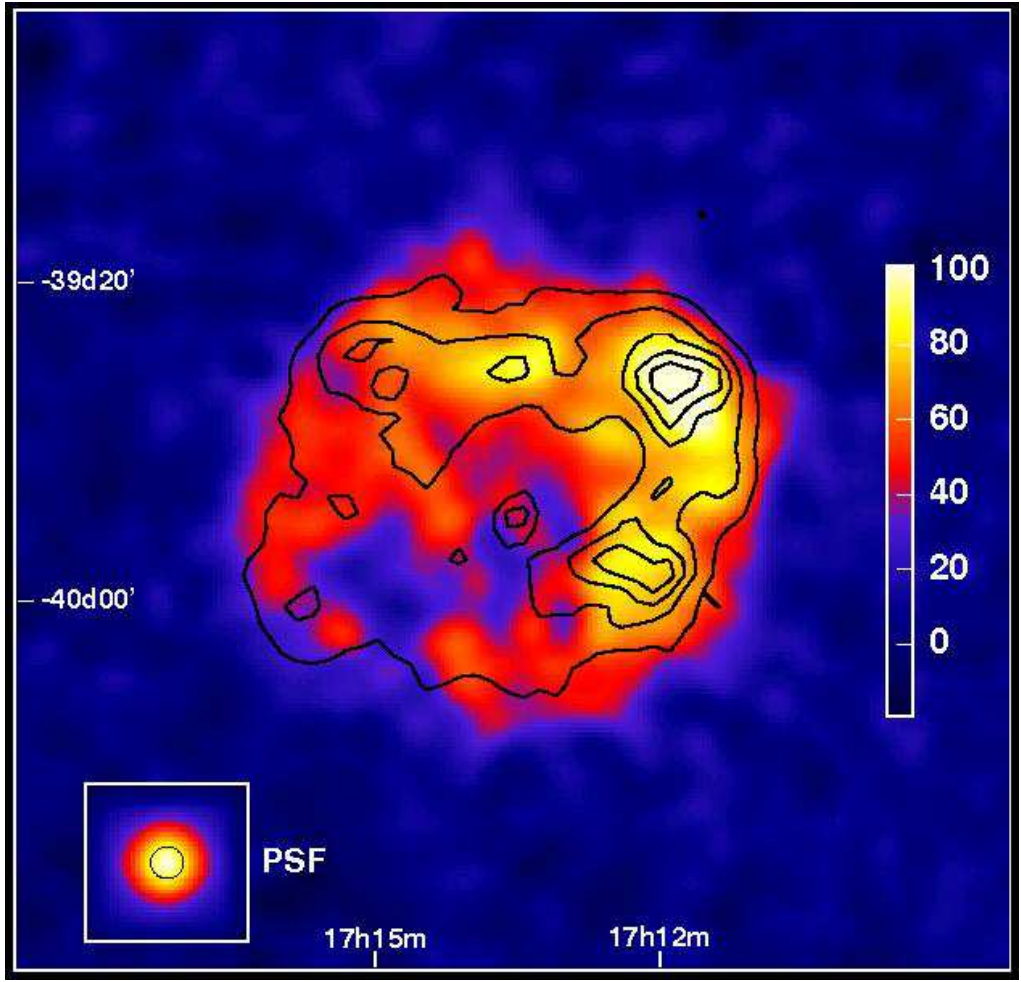}
  \caption{Spatially resolved VHE $\gamma$-ray image of the SNR RX J1713.7-3936
    as obtained by the {\it H.E.S.S.} array. The {\it ASCA} hard X-ray data
    ({\it contour lines}) are shown in addition.}
            \label{fig:RXJ}
\end{figure}

Cas~A is a thoroughly  studied object in all wavelength ranges below
$\gamma$-ray energies and thus amenable to a detailed application of nonlinear
acceleration theory (\cite{Berezhko03b}). In addition, the observations of
narrow filamentary structures in nonthermal X-rays (assumed to be synchrotron
radiation) by \cite[Vink \&  Laming (2003)]{Vink03} can be interpreted to imply
an interior magnetic field strength of $500 \mu$G  (\cite{Berezhko04}),
consistent with the theoretical analysis of the spatially integrated
synchrotron spectrum. This strongly suggests Cas~A as a source of nuclear
Cosmic Rays, independent of all remaining uncertainties in the astronomical 
parameters.

% \begin{figure}
%  \includegraphics{fig07.eps}
%   \caption{VHE $\gamma$-ray image of ``Vela Jr'' as obtained by the {\it
%   H.E.S .S.} array.
%            The {\it contour lines} are smoothed X-ray data from the {\it
%            ROSAT} All
%            Sky Survey, at energies above 1.3 keV.}
%             \label{fig:VelaJr}
% \end{figure}
%

The {\it H.E.S.S.} observations of RX J1713.7-3946 have for the first time
resolved the morphology of an extended object in VHE gamma rays (Fig.5). The
overall shell structure coincides closely in hard X-rays and gamma rays.
Despite the complex structure this is unambiguous proof of the acceleration of
charged particles to energies beyond 100 TeV.  The most recent data show in
addition that below about 10 TeV the photon spectrum can be approximated by a
power law in energy with an index close to 2.0. Such a spectrum is clearly
consistent with diffusive shock acceleration. More detailed studies are in
progress.

RX J0852.0-4622, also called ``Vela Jr'', was, like RX J1713.7-3946, first
detected with the {\it ROSAT} telesope in X-rays (\cite{Aschenbach98}). The
object has a radius of $\sim 1^{\circ}$, twice that of RX J1713.7-3946 and thus
four times the radius of the full Moon.  The TeV $\gamma$-ray morphology
correlates again very well with the X-ray image, and the energy spectrum -- not
yet very precisely determined -- can be fit by a power law with photon index
2.1. In this sense Vela Jr is similar to RX J1713.7-3946. Its apparently
regular spherical shape and high $\gamma$-ray flux, comparable with that of the
Crab Nebula, make it an ideal object for further studies.

\begin{figure}
\includegraphics[width=5.3in,angle=0]{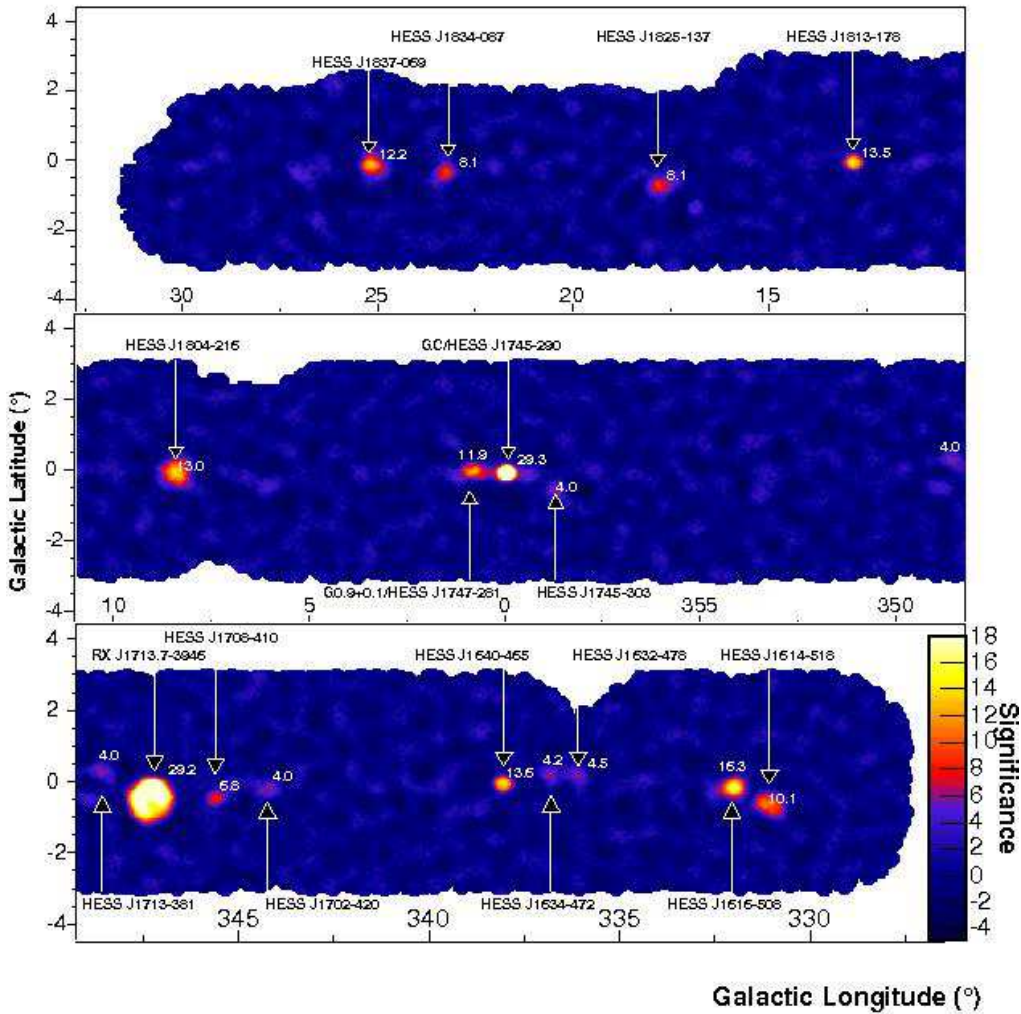}

  \caption{{\it H.E.S.S.} scan of the Galactic Plane between $\pm 30^{\circ}$ 
            in longitude and $\pm3^{\circ}$ in latitude. The sources indicated 
            are detected at a significance level greater than $4 \sigma$.}              \label{fig:Galscan}
\end{figure}

\section{The {\it H.E.S.S.} survey of the Galactic plane}
 
In 2004 {\it H.E.S.S.} has completed the first 230 hour part of a VHE survey of
the Galactic plane, covering $-30^{\circ} < \ell < 30^{\circ}$ in longitude and
$-3^{\circ} < b < 3^{\circ}$ in latitude (\cite{Aharon05g}). The average flux
sensitivity of the survey corresponds to $\sim 3$ percent of the Crab Nebula
(Fig. 6). Including recent re-observations of candidate sources from the
initial survey fourteen previously unknown VHE sources have been found up to
now (\cite{Aharon06}). Most of the sources are still un-identified in other
wavelength ranges.

In this manner for the first time a TeV instrument has investigated not only
source candidates that were well known from other wavelength ranges, but did a
successful  blind search. The search for counterparts in other wavelength
ranges is ongoing. Most known counterpart candidates are young Pulsar Wind
Nebulae or  SNRs. The new sources have triggered a worldwide activity with
satellite X-ray instruments like {\it INTEGRAL, ASCA, XMM} and 
{\it Astro-E2}, and ground-based radio telesopes.

\begin{acknowledgments}
I would like to thank the members of the {\it H.E.S.S.} collaboration for many
discussions on the observational results. I have learnt much from Felix
Aharonian and Okkie de Jager about the physics of PSR
B1259-63, even though I have of course the sole responsibility for 
the arguments in the text.
\end{acknowledgments}

\end{document}